\documentclass[twocolumn]{revtex4}
\usepackage{graphicx}
\usepackage{latexsym}
\def\LSNiO{La$_{2-x}$Sr$_x$NiO$_{4+\delta}$}
\def\BSCCO{Bi$_2$Sr$_2$CaCu$_2$O$_{8+\delta}$}
\def\LNSCO{La$_{1.6-x}$Nd$_{0.4}$Sr$_x$CuO$_{4}$}

\begin{document}
\hyphenation{Ka-pi-tul-nik}

\title{STM Studies of Near-Optimal Doped \BSCCO}

\author{A. Kapitulnik$^{1}$, A. Fang $^1$, C. Howald$^1$ and M. Greven$^{1,2}$ \\
$^1${\it Departments of Applied Physics and Physics, Stanford University,  Stanford, CA 94305}\\
$^2$ {\it Stanford Synchrotron Radiation Laboratory, Stanford, CA 94309}}


\begin{abstract}
In this paper we summarize our STM studies of
the density of electronic states in nearly
optimally doped \BSCCO in zero field. We report on the inhomogeneity of the gap structure, density of states modulations with four-lattice constant period, and coherence peak modulation. 
\end{abstract}

\maketitle
\section{Introduction}
\vspace{-2mm}
Tunneling spectroscopy has been an important tool in the study of
high-temperature superconductors since their  discovery. In the
early days of high-$T_c$ a variety of gap sizes and structures
were found and introduced much controversy into the subject.
However, recent measurements have been more consistent among
groups, revealing a relatively coherent picture of the surface of
high-Tc materials as viewed with STM \cite{kirk,renner1,yazdani,davis1,davis2,renner2,davis3}.  In this paper we review our work on the fine-scale structure of the electronic states at the surface of \BSCCO (BSCCO) as revealed by STM measurements \cite{howald1,howald2,howald3,fang}.  Starting with an analysis of the shape of individual spectra, we further look at the spatial dependence of features such as the size of the gap, coherence peaks and local density of states (LDOS) modulations.  We perform measurements using a  cryogenic STM on  near optimally doped BSCCO ($T_c \sim 86$K) grown by a floating-zone
method.  The samples are cleaved in an UHV
of better than  $1\times 10^{-9}$  torr and then quickly lowered to
the cryogenic section at a temperature of 6-8K.  Most data were taken with
a sample bias of -200mV  and a set point current of -100pA
At each point on the surface, dI/dV spectra was also taken.

\section{Individual Spectral Shape}
\vspace{-2mm}

\begin{figure}[h]
 \includegraphics[width=1.0 \columnwidth]{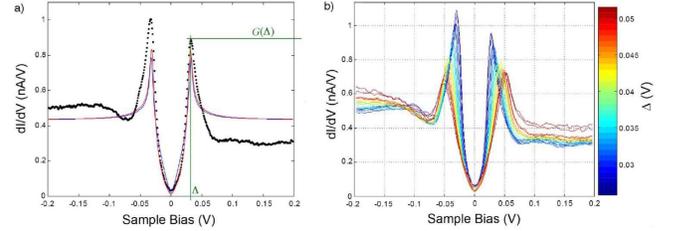}
 \caption{\footnotesize a)  Typical spectrum with fits.  Blue curve shows best d-wave fit, red shows  d-wave fit with angle-dependent Fermi velocity.  b) Series of spectra with coherence peaks for increasing gap sizes. }
\label{specplusfits}
\end{figure}

 A representative spectrum is shown in Fig.~\ref{specplusfits}a.
To analyze the spectrum we use a  d-wave gap formula $\Delta (\theta)=\Delta cos(2\theta)$
with thermal broadening $k_B T$, and smearing $\Gamma$, averaged over all in-plane
{\bf k}-directions with a weight function $g(\theta)$ appropriate for tunneling perpendicular to the CuO$_2$ planes.

\begin{equation}
\footnotesize{ \frac{N_S(E)}{\langle N_N\rangle }= 
\int_{-\infty}^{\infty}\hspace{-2mm}dE' \mathcal{R}e \left\{ \int_0^{2\pi}\frac{g (\theta)}{2\pi}d\theta \frac{E-i\Gamma}{\sqrt{(E-i\Gamma)^2 - \Delta(\theta)^2}} \right\} \frac{df}{dE'}}
\label{eq1}
\end{equation}

\noindent Where $f(E-E')$ is a Fermi function and the Lorenzian broadening is sometimes replaced by a gaussian one with width $\Gamma$. In general a finite $\Gamma$ tends to reduce the coherence peaks. However, as can be seen from Fig.~\ref{specplusfits}b, spectra with small gaps tend to have unusually tall coherence peaks.  Thus, for the fit in Fig.~\ref{specplusfits}a we use  $\Gamma=0$, obtaining $\Delta=32 mV$ and $k_B T = 0.7 mV
\pm 0.3 mV$ (consistent with the measurement temperature of  8 K). 

The fit fails to capture elements of the data in several areas. First, there is always a significant
particle-hole asymmetry in the data. While this probably comes
from the non-flat shape of the normal state DOS, and could be
added to the fit in the form of an asymmetric background, the
d-wave form, even with $\Gamma =0$ can still not fit the excessive height of the
coherence peaks, nor the dip at approximately twice the gap size
which is more pronounced on the negative bias side. The need for a $g(\theta)$ that is different from unity was first proposed by Oda {\it   et al.} \cite{oda} to explain the more rounded shape of the gap at zero bias. The red curve in Fig.~\ref{specplusfits}a shows a fit of this type, using $g(\theta)$ which
varies by a factor of two between the node and the antinode,
yielding $\Delta=31 mV$ and $k_B T = 1.2 mV \pm 0.3 mV$.  This fit
captures the shape of the subgap DOS, though there appears to be
a slight excess of measured states at very low bias, perhaps due
to zero bias anomalies.     This fit accentuates the excess states at the coherence peaks as it fits the full amount of states removed from the gap region, but produces  a  coherence peak that is weaker than the one found experimentally.  This effect is strong at small gaps and become weaker with larger gaps and lower coherence peaks.  While there is no theoretical explanation for this result, this finding may point to a peculiar effect in which in the superconducting state the coherence peak gains states from high energies as well as from the gap region.  The fact that small gaps show larger coherence peaks may point that this effect is stronger in more overdoped samples as will be discussed in section \ref{gdelta}.  For larger gap and lower coherence peaks spectra a finite $\Gamma$ is needed.  A spectrum with a gap size $\Delta = 42$ mV will typically yield $\Gamma \approx 2 -3$ mV if Eqn.~\ref{eq1} is used.  While a precise determination of the  numerical value of the gap is therefore impossible at present,  there is fairly good agreement between the  gap values found by
the fits and the location of the maximum in the $dI/dV$, so we will
use this maximum $G(\Delta)$ as a phenomenological measure of $\Delta$, 
throughout.  For spectra that do not show coherence peaks we will use the edge where the conductance decreases below the background as a measure for the gap \cite{howald1}.

The spectra shown in Fig.~\ref{specplusfits}b are  typical when coherence peaks are present.  There are several observations we can make.  First is that the gap can vary by as much as a factor of 2 within a
scan area.  Second, it is clear that the larger gaps have lower coherence peaks. Also, the energy of the dip feature seems to move with increasing gap size. This dip has been attributed to a strong coupling
effect \cite{dewilde}, regardless, it is apparently related to
superconductivity since its energetic location is scaling with the
gap.  Finally, the negative bias slope increases with gap size,
while the bias asymmetry decreases. Presumably these latter
changes reflect the evolution of the normal (non-superconducting)
density of states. Also note that the asymmetry in the coherence
peak heights changes with the gap size.  The largest gaps have
more weight in the positive bias coherence peak, while the
smaller gaps have more weight in the negative bias peak.

\section{Spatial Variations of the superconducting properties}
\vspace{-2mm}

\begin{figure}[h]
\begin{center}
\includegraphics[width=1.0 \columnwidth]{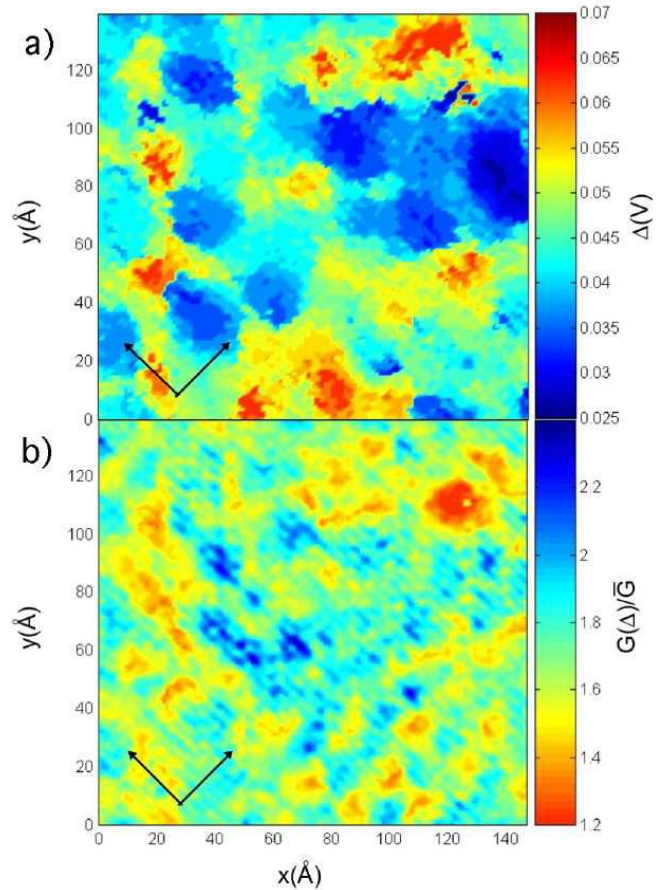}
\end{center}
\vspace{-4mm} \caption{\footnotesize Comparison of a) gap size and b) coherence peak height over a
$140\AA \times 140 \AA$ area. Arrows denote the Cu-O bond direction.} 
\label{variation}
\end{figure}

 Fig.~\ref{variation}a shows a typical map of the size of the gap. Patches about 30 $\AA$ across of varying gap size are clearly visible. While the magnitudes of the largest gaps may
vary between samples (mostly dependent of fraction of gaps with no coherence peaks),  the magnitude of the smallest gaps observed is always $\sim 30$ mV.  The
smallest scale features reflect some variation with atomic
resolution, and the partial near vertical lines show that
there is some correlation between superstructure and the gap.

Spatial variations of the gap on the surface of BSCCO, similar to Fig.~\ref{variation}a were reported by several groups \cite{howald1,pan1,lang}. However, the origin of the inhomogeneities is still not clear. Martin and Balatsky \cite{martin} proposed a phenomenological model where  disorder in the dopant
sites will lead to variations in the carrier density in the plane due to the unusually large screening length of doped Mott insulators. The local doping is therefore the average density of
dopant atoms within a superconducting coherence length lead to patches with differing superconducting properties, particularly the gap size.  Wang {\it et al.} \cite{wang}, used a t-J model  to show  that the poor screening results in an effective, larger than the actual distance,  separation between the CuO$_2$ planes.   Both calculations showed that simultaneous measurements of the spatial extent of the doping  variations and the accompanying gap size variations provides a check on whether the doping inhomogeneities can  account for the gap inhomogeneities.  

\begin{figure}[h]
\begin{center}
\includegraphics[width=1.0 \columnwidth]{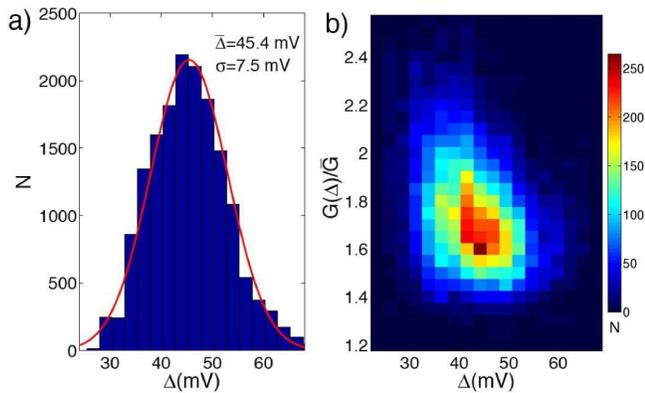}
\end{center}
\vspace{-4mm} \caption{\footnotesize a) Histogram of gap size as dervived from Fig.~\ref{variation}a.  b) Coherence peak conductance vs. gap size derived from Fig.~\ref{variation}} 
\label{histog}
\end{figure}

Fig.~\ref{histog}  shows a histogram of the gap size over
the area shown in Fig.~\ref{variation}a  with  the best gaussian fit to this histogram.
The width of the gap distribution is about 7.5 mV. The model of Martin and
Balatsky, using  the experimental value of the coherence
length, $\sim$15$\AA$, and  a gap that is approximately linear in doping with a
coefficient of 0.3V/carrier, yields a standard deviation of the gap size of $\sim$30mV . This is much larger than we found experimentally. The model of Wang {\it et al.} uses a different gap versus
doping dependence but  yields similar, though slightly smaller, discrepancies.   The above analysis may point that other models that are more qualitative and based on electronic phase separation may be the true explanation for the gap inhomogeneities phenomenon as was previously discussed by Howald {\it et al.} \cite{howald1}. 

The spatial variation of the coherence peak height
differs somewhat from the spatial variation of the gap.
Fig.~\ref{variation}b shows the variation of the
coherence peak height in the same area of Fig.~\ref{variation}a.  While
there is a definite correlation between gap size and coherence
peak height,  there are several differences. First,
the peak height varies considerably on the atomic scale, while
the peak position does not.  Second, the peak position more
clearly exhibits the granular structure: except for the atomic
scale variations, the variation in the peak height is smoother. Finally, it is clear from Fig.~\ref{variation}b that the peak-height shows an ordered structure, especially if one considers the lower-right corner of the figure. We discuss this phenomenon in  section \ref{gdelta}

\section{Periodic structures in the LDOS: General Considerations}
\vspace{-2mm}

The discovery \cite{hayd92,chen93,tran94a} of stripe
order in {\LSNiO}  and soon after in {\LNSCO} \cite{tran95a} added considerable credibility
to the suggestion that charge ordered states form an important bridge between the
Mott insulator, and the more metallic state at heavy doping.
STM however is a static probe and thus cannot detect any structure associated
with fluctuating order unless something pins it.  Indeed the inhomogeneities discussed above and other point defects are a natural source for pinning. As a result charge order will be visible to STM in the form of LDOS modulations.  However, defects can also create other effects that need to be understood before a static charge modulation explanation is invoked.

Defects in simple metals will also cause LDOS modulations.   These ``Friedel oscillations"  \cite{friedel}   are related  to Fermi-surface-derived  non-analyticities in the susceptibility,
$\chi(\bf k)$.   A generalized form of these oscillations  can occur
in more diverse systems in which the relevant structure in $\chi(\bf k)$ is not directly related to  any feature of a Fermi surface.   In particular, if the system is proximate to a charge modulated state, such as a stripe state, the values of $\bf k=\bf q$ at which $\chi$ has maxima will reflect the pattern of spatial
symmetry breaking of the ordered state, but $\chi(\bf k)$ will respect the
full point-group symmetry of the crystal (unless the liquid state  is a nematic \cite{kivelson}).  So, the generalized Friedel oscillations around a point impurity in a stripe-liquid phase will inevitably form a checkerboard pattern \cite{polkovnikov,kivelson}.

There is another form of spatial modulation of the density of states, one with a period which disperses as a function of the probe energy.  This latter effect, which was first demonstrated by Crommie {\it et al.} \cite{crom93}, is produced by the elastic scattering of quasiparticles of given energy off an impurity.  The resulting interference between scattered waves leads to variations of the local density of states at wave vectors $\bf{q}=\bf k-\bf k^{\prime}$, where $\bf k$ and $\bf k^{\prime}$ are the wave-vectors of states with
energy $E_{\bf k}=\epsilon_{\bf k}=\epsilon_{\bf k^{\prime}}$, as determined by the band structure,
$\epsilon_{\bf k}$.  Generalized versions of these oscillations can occur even when there are no well defined quasiparticles, so long as there are some elementary excitations of the system with a well-defined dispersion relation. However, quasiparticle scattering interference will take place at a certain energy and wave-vector \emph{only} if there are available states.  Concentrating on $q_{\pi-0} = \frac{1}{4} (2\pi /a_0)$, 
 It is easy to see that the quasiparticle interference picture cannot produce such
peaks at low energies \cite{kivelson,wanglee}. Assuming
quasiparticles with {\bf k}-dependent energy: $E_{\bf k}=\sqrt{\Delta_{\bf k}^2 +
\epsilon_{\bf k}^2}$, where $\Delta_{\bf k}={{\Delta_0} \over {2}}[cos(k_xa_0)
- cos(k_ya_0)]$ is the d-wave superconducting gap.  For the above $q$ vector we can take ${\bf q}=(2\pi/4a_0, 0)$. In that case ${\bf k}=(-\pi/4a_0, k_y)$, and ${\bf k^\prime}=(\pi/4a_0,
k_y)$.  If we extract $k_y$ directly from the ARPES data
\cite{bogdanov1,kaminski}, we find $k_y \approx 0.6 (\pi/a_0)$.
For $\Delta_0 \sim 30$ meV \cite{loeser,ding1,fedorov}, which is
also the minimum gap found on this type of samples using STM,
\cite{howald1,lang}  this gives an estimate of the lowest energy
for quasiparticle scattering at this wavevector of approximately
$\Delta_0/2 \sim 15$ meV.  This energy cutoff  could differ a little, depending  on the exact details of the
band-structure, but this analysis certainly excludes the energies
around zero bias.

However, the above sources for LDOS modulations should not appear exclusively.  In fact, charge order and quasiparticle scattering are likely to coexist in a system with well defined quasiparticles. \cite{vojta,podolsky,kivelson} The relationship among these effects will be discussed below.

\section{Periodic structures in the LDOS: Experimental result }
\vspace{-2mm}

The essence of the above discussion is that in STM studies of cuprates we would expect stripe or checkerboard correlations to make an appearance as generalized Friedel oscillations, while
quasiparticle-like interference is a distinct phenomenon that could also
be present.  The observation of a checkerboard pattern with a $\sim 4a_0$ period
about vortex cores in BSCCO  \cite{hoffman1} has been a possible
evidence for pinned charge stripes.  However, it remained of great importance to find evidence of charge modulation in BSCCO with no applied field.  

\begin{figure}[h]
\begin{center}
\includegraphics[width=1.0 \columnwidth]{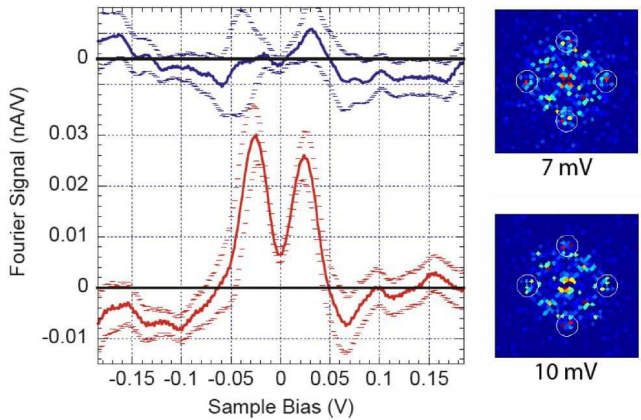}
\end{center}
\vspace{-4mm} \caption{   \footnotesize Fourier transform at  q=(2$\pi /4a_0$)(0,$\pm$1), the location of the peaks  as a function of sample bias. The red and blue
traces correspond to the real and imaginary parts, respectively. Right: Two Fourier transforms maps at low energies showing peaks at $q_{\pi-0} \sim [0.25\pm0.03](2\pi/a_0)$ and $q_{0-\pi} \sim [0.25\pm0.03](2\pi/a_0)$.  Circles denote the main contribution to the peaks.} 
\label{ft}
\end{figure}

Indeed, Howald {\it et al.}\cite{howald2} shortly afterwards reported this same effect in zero field on similarly doped BSCCO crystals fabricated without intentional substitution of  impurities. The observed modulation with ordering wave vector $q_{\pi-0} \sim [0.25\pm0.03](2\pi/a_0)$ was found at all energies,  exhibiting features characteristic of a two-dimensional system of line objects.  Moreover, Howald {\it et al.} showed that the LDOS modulation manifests itself, for both positive and negative bias, as a shift of states from above to below the superconducting gap. The fact that a single energy scale (i.e. the gap) appears for both superconductivity and these modulations suggests that these two effects are closely related. The summary of the results of Howald {\it et al.} is shown in Fig.~\ref{ft}, emphasizing the fact that peaks in the Fourier transform at $q_{\pi-0} \sim [0.25\pm0.03](2\pi/a_0)$ are found even at low energies where quasiparticle scattering interference should not produce a signal.  Fig.~\ref{ft} was reproduced theoretically by Podolsky {\it et al.} \cite{podolsky}.

Subsequent studies at zero field \cite{hoffman2,mcelroy} measured the dispersion of the strongest Fourier peak along the $\pi-0$ (i.e. Cu-O) direction.  They asserted that it was consistent with what is expected from quasiparticle scattering interference. \cite{wanglee}  In general, their data showed good agreement with photoemission results  (i.e. band structure results \cite{damascelli}) at large bias, but unlike photoemission results, did not continue to disperse below $\sim$ 15 mV.

\begin{figure}[h]
\begin{center}
\includegraphics[width=1.0 \columnwidth]{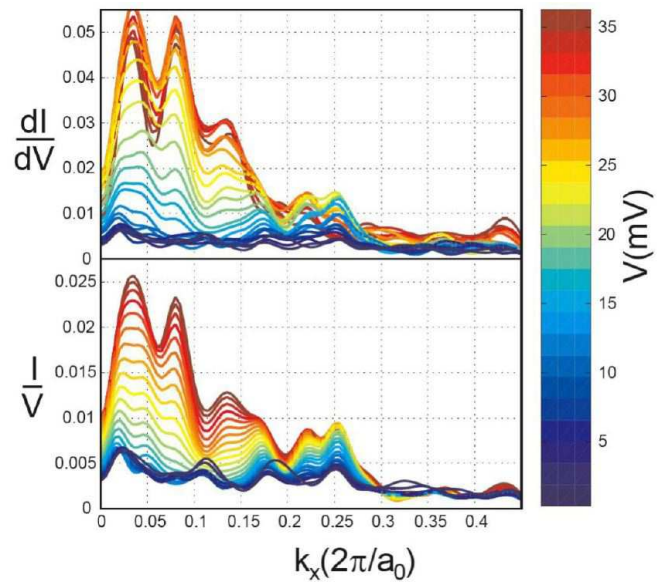}
\end{center}
\vspace{-4mm} \caption{\footnotesize Line scans as a function of $k_x$ along the $(0,0)$ to
$(\pi,0)$ direction, and as a function of energy (color scale).
Top panel shows the LDOS (dI/dV), and bottom panel the integrated
LDOS  (I) up to the given energy.} 
\label{integ}
\end{figure}

To resolve the discrepancy Kivelson {\it et al.} \cite{kivelson} proposed several procedures which allowed Howald {\it et al.} to separate the two effects. First,   integration of the LDOS over a wide range of energies reduces the influence of any random or dispersing features such as quasiparticle scattering interference, while at the same time it enhances features that do not disperse.  Fig.~\ref{integ} shows the result of such a procedure. 

A complementary approach to separate pure charge modulation effects is to look for an interaction of the charge modulation with the superconducting order parameter.  We claim that the periodic structure observed in $G(\Delta)$ in Fig.~\ref{variation}b is exactly this effect. We discuss this procedure in the next section.

\section{Modulation in $G(\Delta)$}
\label{gdelta}
\vspace{-2mm}

\begin{figure}[h]
\begin{center}
\includegraphics[width=1.0 \columnwidth]{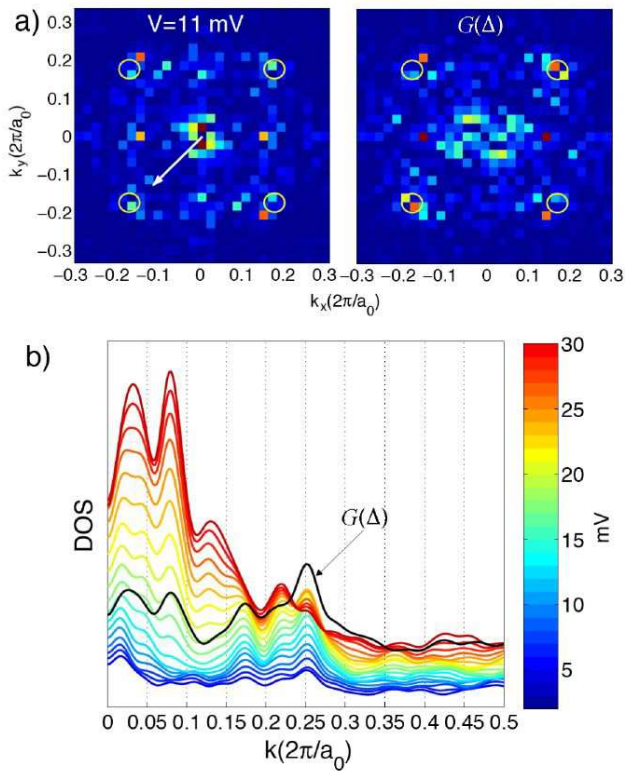}
\vspace{-4mm} \caption{ \footnotesize a) FFT of LDOS at 10 mV and of LDOS  taken at the coherence peak maximum with 65 mV normalization (see text.)  b) Dispersion relation of the charge modulation periodicity.  Black 
line is normalized coherence peak maxima.} 
\label{peakdos}
\end{center}
\end{figure}

Coming back to Fig.~\ref{variation}, we compare the coherence peak heights and the gap size. The data for this figure was taken with a setpoint voltage of 65 mV.  The pronounced enhancement of the modulation signal with this setpoint voltage led us to a novel procedure to normalize (i.e. divide) 
the individual spectra by the current at $+65$ mV.  A detailed description of this procedure is given in Fang {\it et al.} \cite{fang}.  Fig.~\ref{peakdos} shows a representative Fourier transform of low bias together with a Fourier transform of the peak in the $dI/dV$ (i.e. $G(\Delta)$) for a $160\AA \times 160\AA$ area sample \cite{howald2}.  A clear correspondence of the peaks in the $0-\pi$ direction is found when comparing these two maps. This is a striking result since $\Delta$ is in the range of 30 - 60 mV which according to band structure is the strongly dispersive region of quasiparticles scattering interference. 

Fig.~\ref{peakdos} also shows a line scan in the $0-\pi$ direction together with $G(\Delta)$ for the same sample.  For all our samples, our Fourier analysis from low energies up to the smallest gap sizes (where the noise from inhomogeneities overwhelms our signal) supports the picture presented earlier of a non- or weakly  dispersive feature in the region $q= 0.22 (2\pi/a_0) - 0.25 (2\pi/a_0)$ in addition to dispersive features at a lower $k$-vector.
We find that the large amplitude of the lower $k$-vector features swamp out the non-dispersive feature at higher energies, but become relatively weak at lower energies.
The additional line scan for the Fourier transform of coherence peak heights can be
seen as a way to remove the effects of gap size inhomogeneities to reveal that a structure
at $q\approx 0.25 (2\pi/a_0)$ still exists at higher energies.  It makes a similar
point as the spatial maps of the coherence peaks, namely, that by selectively
sampling from the (higher) energies related to superconductivity, the
low energy features reappear.

By comparing Figs.~\ref{variation}a and \ref{variation}b,
one can see that the amplitude  of the coherence peak DOS modulations is larger in the regions of large gap.   In contrast, regions of small gap show modulations of reduced amplitude.  
Since there are only a few modulation crests and troughs within a particular
region of large or small gap, this effect is difficult to quantify, although
it can most easily be seen by following the regions of largest gap.   We note that the regions of large gap with low coherence peaks generally resemble slightly underdoped samples\cite{howald1}.  On the other hand, the modulation is suppressed and the coherence peak heights are more uniform in regions of small gap and tall coherence peaks.  Gaps  in these regions are more similar to gaps found in overdoped samples \cite{ozyuzer}; this is consistent with the notion that beyond optimal doping a more homogeneous  charge density exists closer to a Fermi liquid state.
 
Such an observation does not necessarily point out a competition between charge-density modulation and superconductivity, but rather reinforces the idea that the two effects coexist at and below optimal doping.  One possible interpretation is that the fluctuating stripe/checkerboard phase exists  in all the regions below optimal doping, and as one moves further 
into overdoping (i.e. into regions of small gap) the modulations become diminished.
 Our observations therefore complement those of Vershinin {\it et al.}\cite{vershinin} who found similar patterns in the pseudogap regime of slightly underdoped \BSCCO.  
As noted by Kivelson {\it et al.} \cite{kivelson}, the effect of quasiparticle scattering interference should disappear at temperatures above $T_c$, revealing the underlying order.
For our measurements at low temperature, in regions of very large gap with weak coherence peaks (similar to the pseudogap), charge ordering is indeed visible.  The two results therefore suggest that in the absence or suppression of superconductivity, charge-ordering may be the preferred phase.

\section{Conclusions}
\vspace{-2mm}

In this paper we discussed the evolution of structures in the LDOS on the microscopic scale. We first analyzed the individual spectra pointing to anomalies in their shape and discussing the determination of the size of the superconducting gap.  Maps of the gap lead to an inhomogeneous pattern that cannot be explained by simple disorder in doping and thus may point to intrinsic electronic phase separation. The inhomogeneities also reveal ordered checkerboard patterns in the LDOS with a period close to four lattice constants. Studying the interplay between the two effects we suggest that both have a common origin.\\

\noindent {\bf Acknowledgements:} We thank  Steven Kivelson for many helpful discussions. Work supported by the Department of Energy grant DE-FG03-01ER45925.

\end{document}